\documentclass[openacc]{rstransa}

\begin{document}
\title{The Galerkin-truncated Burgers equation: Crossover from inviscid-thermalised 
to Kardar-Parisi-Zhang scaling}

\author{
C. Cartes$^{1}$, E. Tirapegui$^{2}$, R. Pandit$^{3}$ and M. Brachet$^{4}$}

\address{$^{1}$ Complex Systems Group,
Facultad de Ingenier\'\i a y Ciencias Aplicadas,
Universidad de los Andes,
Santiago, Chile.\\
$^{2}$ Departamento de F\'\i sica, 
Universidad de Chile, 
Santiago, Chile.\\
$^{3}$ Centre for Condensed Matter Theory, Department of Physics, 
Indian Institute of Science, Bangalore 560012, India.\\
$^{4}$ Laboratoire de Physique de l'\'Ecole Normale Sup\'erieure, ENS, Universit\'e PSL, CNRS, Sorbonne Universit\'e, Universit\'e de Paris, F-75005 Paris, France.
}

\subject{Fluid Dynamics, Statistical Mechanics}


\corres{Marc Brachet\\
\email{marc-etienne.brachet@ens.fr}}

\begin{abstract}
The one-dimensional ($1D$) Galerkin-truncated Burgers equation, with both dissipation and noise terms included, is studied using spectral methods. 
When the truncation-scale Reynolds number $R_{\rm min}$ is varied, from very small values to order $1$ values, the scale-dependent
correlation time  $\tau(k)$ is shown to follow the expected crossover from the short-distance $\tau(k) \sim k^{-2}$ Edwards-Wilkinson scaling to the 
universal long-distance Kardar-Parisi-Zhang scaling $\tau(k) \sim k^{-3/2}$.
In the inviscid limit: $R_{\rm min}\to \infty$, we show that the system displays {\it another} crossover to the Galerkin-truncated inviscid-Burgers regime that 
admits thermalised solutions with $\tau(k) \sim k^{-1}$.  The scaling form of the time-correlation functions are shown to follow the known 
analytical laws and the skewness and excess kurtosis of the interface increments distributions are characterised.
\end{abstract}


\begin{fmtext}
\section{Introduction}
Galerkin-truncated hydrodynamical systems, which
retain only a finite number of Fourier modes, 
have been studied actively in fluid
mechanics \cite{LEE:1952p4100,hopf1952statistical,KRAICHNAN:1955p3039,KRAICHNARH:1973p2909,OrszagHouches}.
\end{fmtext}
\maketitle
In his pioneering work \cite{LEE:1952p4100} of 1952, T.D. Lee showed that these truncated systems satisfy Liouville's theorem
and that, assuming ergodicity, there is energy equipartition among the spectral modes.
Later, Kraichnan \cite{KRAICHNARH:1973p2909} proposed a different approach for these {\it absolute equilibrium} states
by considering that the complex amplitudes of the Fourier modes followed a canonical distribution that is controlled by the mean values 
of the invariants of the system.
The Galerkin-truncated hydrodynamical
system that has been investigated most extensively is the time-reversible Euler
equation for a classical, ideal fluid \cite{Cichowlas:2005p1852,Krstulovic:2009p1876}, which can be studied efficiently, in a
spatially periodic domain, by the Fourier pseudospectral
method~\cite{Got-Ors,Canuto-Hussaini}.
Absolute-equilibrium solutions have also been examined in
a variety of hydrodynamical systems including compressible flows \cite{Krstulovic:2009IJBC}, 
the Gross-Pitaevskii equation in
both three (3D) and two (2D) dimensions~\cite{Krstulovic2011a,Shukla2013}, and the
Euler equation and ideal magnetohydrodynamics (MHD) in 2D~\cite{Pouquet2011}.

These results on thermalization in hydrodynamical systems are known well in the
fluid-dynamics community, but less known, than they deserve to be, in the area
of nonequilibrium statistical mechanics. To bridge this gap between these
related fields, we carry out a systematic study of the relaxation to absolute equilibrium in
the $1D$ inviscid Burgers equation, perhaps the simplest hydrodynamical system
demonstrating absolute equilibration. The initial stages of the thermalization are known to involve 
the formation of oscillatory structures, which have been named
\textit{tygers}~\cite{Ray2011,Ray2014,Ray2015}. 
The typical relaxation time near the absolute equilibrium can be studied
conveniently via the scale-dependent correlation time $\tau(k)$, which can be
computed from the time-dependent correlation function.
It is known to scale like the eddy turnover time, that is
as $\tau(k) \sim k^{-1}$ \cite{Majda2000,Majda2002,Cichowlas_Thesis}.
Note that the same $ k^{-1}$-scaling law is known to take place in the 
truncated $3D$ Euler equation \cite{Cichowlas:2005p1852}.

Adding noise and dissipation terms to the $1D$ inviscid Burgers equation
(see equation (4.14) of reference \cite{FNS}) 
transforms it into the Kardar-Parisi-Zhang (KPZ) equation \cite{KPZ,thhzhang1995,thhkat2015,quastel2015}
that is well known in nonequilibrium statistical mechanics. 
The KPZ equation admits the same exact
equilibrium probability distribution as the inviscid Burgers equation 
\footnote{in only $1D$. In dimensions greater than $1$, the inviscid Burgers equations does not conserve the energy (see \textit{e.g.} Ref. \cite{FNS}).}. 
However, the $1D$
KPZ correlation time around equilibrium is known to have a $k^{-3/2}$
scaling. The different time-correlation scalings $k^{-1}$ and $k^{-3/2}$ around
the same equilibrium for the inviscid truncated Burgers equation and the KPZ equation are the main
motivation for the present work.  Note that there is
also a third (trivially linear) viscous type of scaling known as the 
Edwards-Wilkinson\cite{EW1982} (EW) scaling $k^{-2}$
that arises when the nonlinear term is negligible.
In the following, we will 
characterise the crossover behaviour between these different regimes in terms of the Reynolds number 
estimated at the truncation scale.

The remainder of this paper is organised as follows. 
Section \ref{sec:sysdef} contains the system's definitions,
with special attention given to spectral truncation, conserved quantities and
stationary probabilities.
Section \ref{sec:numres} is devoted to our numerical results:
after defining the algorithms and physical parameters,
the scalings of the correlation times
and the distributions of the interface increments are characterized.
Finally, our conclusions are given in Section \ref{sec:conc}.
\section{System definitions} \label{sec:sysdef}
We consider the randomly forced, generalized $1D$ Burgers equation that is 
defined by the following stochastic partial differential equation for the velocity 
field $u(x,t)$ (see, e.g., Refs.~\cite{FNS,KPZ,thhzhang1995,Halpin-Healy2014,thhkat2015,frischbook,Frisch2001,BecKhaninPhysRep}):
\begin{equation}
\partial_t u + \lambda u \partial_x u=\nu \partial_{xx} u + \sqrt{D} \partial_x f ,\label{eq:Burgers}
\end{equation}
where $\lambda$ is the coefficient of the nonlinear term, $\nu$ is the kinematic
viscosity, $D$ a diffusion coefficient and $f$ is a zero-mean, Gaussian force with variance 
\begin{equation}
\langle f(x,t) f(x',t')\rangle= 2 \pi \delta(x-x') \delta(t-t').
\label{eq:KPZ2} 
\end{equation}
If we define $u \equiv \partial_x h$, i.e.,
\begin{equation}
h(x,t)=\int_0^x u(y,t) dy
\end{equation}
we obtain the Kardar-Parisi-Zhang (KPZ) equation
\begin{equation}
\partial_t h + \frac{\lambda}{2}(\partial_x h)^2=\nu \partial_{xx} h +  \sqrt{D} f .\label{eq:KPZ1} 
\end{equation}
We contrast the following three cases in our study: (i) The deterministic,
inviscid, $1D$ Burgers equation, with $\lambda = 1$, $\nu = 0$, and $D = 0$;
(ii) the Edwards-Wilkinson (EW) equation, with $\lambda = 0$, $\nu > 0$, and $D >
0$; and (iii) the KPZ equation, with $\lambda > 0$,  $\nu > 0$, and $D > 0$.

\subsection{Spectral truncation and conserved quantities}
Henceforth, we consider $2\pi$-periodic 
boundary conditions in $x$.  
We introduce the Fourier representation
\begin{equation}
	u(x,t)=\sum_{k=-\infty}^\infty \hat{u}(k,t) \exp(i k x),
\end{equation}
where the caret denotes a spatial Fourier transform, $u(x,t) \in \mathbb{R} $, 
so $ \hat{u}(-k,t)=\overline{\hat{u}(k,t)}$; complex conjugation is indicated
by the overline.
Using
\begin{equation}
\frac{u^2(x,t)}{2}=\frac{1}{2}\sum_{n,p=-\infty}^\infty \hat{u}_{n-p}(t) \hat{u}_{p}(t)e^{i n x},
\end{equation}
the unforced and inviscid Burgers equation (Eq. \eqref{eq:Burgers} with $\nu=0$, $\lambda=1$ 
and $D=0$) can be written as
\begin{equation}
\partial_t \hat{u}(k,t) = -\frac{i k}{2}\sum_{p=-\infty}^\infty \hat{u}_{k-p}(t) \hat{u}_{p}(t),\label{eq:Burgers2Spec}
\end{equation}
which conserves 
the total energy 
\begin{eqnarray}
E &=& \frac{1}{2 \pi}\int_0^{2 \pi}\frac{u(x,t)^2}{2} dx \\ \nonumber
&=& \frac{1}{2}\sum_{k=-\infty}^\infty |\hat{u}(k,t)|^2.\label{eq:Burgers2Ener}
\end{eqnarray}
Note that integrating by parts the nonlinear term in \eqref{eq:Burgers} shows that the integrals
$I_n(t) = \int_0^{2 \pi}u(x,t)^n dx$ are all conserved by the inviscid dynamics \eqref{eq:Burgers2Spec}
(the energy corresponding to the case $n=2$).

Let us now spectrally truncate (or Galerkin truncate) this system.
To do this, we need to 
\textit{enforce} that, for $k>k_{\rm max}$, $\hat{u}(k,t)=0$ and $\partial_t \hat{u}(k,t) = 0$.
To wit, we introduce the Galerkin projector $\mathcal{P}$ that reads in Fourier space 
\begin{equation}
\mathcal{P}_{\rm G} [ {\hat u}_{k}]=\theta(k_{\rm max}-|k|){\hat u}_{k}, \label{eq:TruncDef}
\end{equation}
where $\theta(k)=1$, if $k\leq k_{\rm max}$ and $\theta(k)=0$, if $k > k_{\rm max}$.
Galerkin truncation amounts to the replacements $u:=\mathcal{P}_{\rm G} [u]$, $ u \partial_x u:= \mathcal{P}_{\rm G} [u \partial_x u]$ and $f:=\mathcal{P}_{\rm G} [f]$ in equation \eqref{eq:Burgers}, thus reducing \eqref{eq:Burgers2Spec} to a finite number of ordinary differential equations.

The Galerkin-truncated version of \eqref{eq:Burgers2Spec} thus reads, for $-k_{\rm max}\leq k \leq k_{\rm max}$,
\begin{equation}
\partial_t \hat{u}(k,t) = -\frac{i k}{2}\sum_{\sup(-k_{\rm max},k-k_{\rm max})}^{\inf(k_{\rm max},k+k_{\rm max})} \hat{u}_{k-p}(t) \hat{u}_{p}(t),\label{eq:BurgersTrunc}
\end{equation}
or, in a more symmetrical form,
\begin{equation}
\partial_t \hat{u}(k,t) = -\frac{i k}{2}\sum_{p,q} 
\delta_{k,p+q} \theta(k_{\rm max}-|p|) \theta(k_{\rm max}-|q|) \hat{u}_{p}(t) \hat{u}_{q}(t), \label{eq:NLTrunc2}
\end{equation}
where $\delta$ denotes the Kronecker symbol.
Thus, the nonlinear truncated term explicitly reads:
\begin{equation}
\mathcal{N}_k(\hat{u})= -\frac{i k}{2}\sum_{p,q} 
\delta_{k,p+q} \theta(k_{\rm max}-|k|)\theta(k_{\rm max}-|p|) \theta(k_{\rm max}-|q|) \hat{u}_{p} \hat{u}_{q}. \label{eq:NLTrunc3}
\end{equation}

It is straightforward to check out that the nonlinear term \eqref{eq:NLTrunc3} verifies the following relations
\begin{eqnarray}
0&=&\mathcal{N}_0(\hat{u}),\nonumber \\
0&=&\sum_{k} \hat{u}_{-k} \mathcal{N}_k(\hat{u}),\nonumber \\
0&=&\sum_{k,p,q} \delta_{-k,p+q} \theta(k_{\rm max}-|p|) \theta(k_{\rm max}-|q|) \hat{u}_{p} \hat{u}_{q}\mathcal{N}_k(\hat{u}).
\end{eqnarray}
Thus, three conservations laws survive the Galerkin truncation and
\begin{eqnarray}
P&=&\hat u_0, \nonumber \\
E &=&\frac{1}{2}\sum_{k=-k_{\rm max}}^{k_{\rm max}} |\hat{u}(k,t)|^2, \rm{and} \nonumber \\
H&=&\sum_{k,p,q} \delta_{-k,p+q} \theta(k_{\rm max}-|k|)\theta(k_{\rm max}-|p|) \theta(k_{\rm max}-|p|) \theta(k_{\rm max}-|q|) \hat{u}_{k}\hat{u}_{p} \hat{u}_{q}
\end{eqnarray}
are \textit{still} conserved after truncation.

The conserved quantities $P$ and $E$ are, respectively, the momentum and the energy of the system.
The third surviving conserved quantity $H$ can be used to provide an explicit Hamiltonian formulation of the truncated system.
It is known to play a role in the thermalization dynamics only for very special choices of the initial conditions \cite{Majda2003}.  

In our direct numerical simulations we use a standard Fourier pseudospectral method, 
with dealiasing performed by the $2/3$ rule. Clearly, such a pseudospectral method is
identical to a spectral Galerkin method  (see, e.g., Ref.~\cite{Got-Ors}). 
We use $N$ collocation points and spectral truncation is 
performed for $k > k_{\rm max}=[N/3]$, where $[\cdot]$ denotes the 
integer part. Note that with this choice of dealiasing, the third conserved quantity $H$ must be evaluated as
$\mathcal{P}_{\rm G}[u \mathcal{P}_{\rm G}[u^2]]$.
If one instead insists, as done in references \cite{Majda2000,Majda2002,Majda2003}, to evaluate it 
simply as $\mathcal{P}_{\rm G}[u^3]$
then the truncation must be performed for $k \ge k_{\rm max}=[N/4]$. Both truncated system
and conserved quantities are identical (when $H$ is evaluated correctly). 
Therefore, here we will use the $2/3$ scheme that allows us to use more modes for a given resolution.

\subsection{Stationary probability}
The nonlinear truncated term \eqref{eq:NLTrunc3} verifies the Liouville property
\begin{equation}
\sum_k \frac {\partial \mathcal{N}_k(\hat{u})}{\partial \hat{u}_k}=0.
\end{equation}
In the case of absolute equilibrium of the
deterministic, inviscid, $1D$ Burgers equation
truncated system, a standard argument (see {\it e.g.}
Refs. \cite{LEE:1952p4100,KRAICHNARH:1973p2909,OrszagHouches}) is that the
microcanonical distribution 
\begin{equation}
P_{\rm mc}[u]=Z_{\rm mc}^{-1}\delta(E-u),\label{eq:Pmicro}
\end{equation}
when the number of degrees of
freedom $2 k_{\rm max}+1$ is large enough,
can be well approximated by the canonical
distribution 
\begin{equation}
P_{\rm sta}[u]=Z_{\rm c}^{-1} e^{-\beta E},\label{eq:pstationary}
\end{equation}
where $Z_{\rm mc}$ and $Z_{\rm c}$ denote normalization factors.

A direct way to proceed is to introduce the Liouville equation
for the probability $\mathbb{P}\left[\{ \hat{u}_{k}, \hat{u}_{k}^*\}_{0\leq k \leq k_{\rm max}}\right]$,
\begin{equation}
\frac {\partial \mathbb{P}}{\partial t}=\sum_{0\leq k \leq k_{\rm max}}\frac{\partial}{\partial  \hat{u}_{k}}\left[ -\mathcal{N}_k(\hat{u}) \mathbb{P} \right]+c.c\, ,\label{eq:Liou}
\end{equation}
where $\hat{u}_{k}^*= \hat{u}_{-k}$ is considered as an independent variable and $c.c$ denotes complex conjugation. 

It follows directly from energy conservation that \eqref{eq:Liou} admits \eqref{eq:pstationary} as a stationary solution.

Note that the stationary distribution \eqref{eq:pstationary} is a white noise in space for $u(x)$ and thus a Brownian process for $h(x)$.

In both the EW ($\lambda = 0$, $\nu > 0$ and $D >0$) 
and the KPZ cases ($\lambda > 0$,  $\nu > 0$ and $D > 0$)
the probability distribution $\mathbb{P}$ of the stochastic process defined by Eqs.
(\ref{eq:Burgers}-\ref{eq:KPZ2}) and the spectral truncation \eqref{eq:TruncDef} 
can be shown to obey the following Fokker-Planck equation \cite{Van-Kampen-Chem,LivreVertET}
\begin{equation}
\frac {\partial \mathbb{P}}{\partial t}=\sum_{0\leq k \leq k_{\rm max}}\frac{\partial}{\partial  \hat{u}_{k}}\left[-(\lambda \mathcal{N}_k(\hat{u})-\nu k^2 \hat{u_k})\mathbb{P}+ D k^2 \frac{\partial \mathbb{P}}{\partial  \hat{u}_{k}^*}\right]+c.c\,.\label{eq:FP}
\end{equation}

Let us remark that
\eqref{eq:pstationary} is also a stationary solution of \eqref{eq:FP}. 
Indeed, the nonlinear term in the Fokker-Planck equation can be treated exactly like its counterpart in the Liouville equation \eqref{eq:Liou}; 
the remaining terms also cancel for the stationary distribution \eqref{eq:pstationary}, 
because, in equilibrium, we must have $\nu k^2 \hat{u}_k - \beta D k^2 \hat{u}_k = 0$, whence we get
\begin{equation}
D=\frac {\nu}{\beta}.\label{Eq:Diffu}
\end{equation}

If we define the {\it r.m.s.} velocity $u_{rms}$ (see Eqs. \eqref{eq:Burgers2Ener} and \eqref{eq:pstationary})
\begin{equation}
 <E>=\frac {u_{rms}^2}{2}=\frac{k_{\rm max}+1}{\beta},
\end{equation}
we find that
\begin{equation}
\beta=\frac{2(k_{\rm max}+1)}{u_{rms}^2}:
\end{equation}
\begin{equation}
D=\frac {\nu u_{rms}^2}{2(k_{\rm max}+1)}.\label{eq:Diffufinal}
\end{equation}

As the equilibrium probability is determined, 
we now focus on the time-correlation
functions 
\begin{equation}
\Gamma(k,\tau)=\langle \overline{\hat{u}_k(t)}\hat{u}_k(t+\tau)\rangle_t. \label{eq:correl}
\end{equation}
In the KPZ case, with the Fokker-Planck equation \eqref{eq:FP}, 
it is well known \cite{FNS,KPZ} that the existence of a fluctuation dissipation
theorem ensures that the associated response function
has the same characteristic time-scale as the equilibrium time correlation function.
The same fluctuation-dissipation relation (with statistical averaging over initial conditions) 
\cite{KRA59} also applies in the inviscid noiseless case \eqref{eq:Liou}.

\section{Numerical results} \label{sec:numres}
\subsection{Algorithms}

We use standard pseudo-spectral Fourier methods. FFTs are performed on $N$
points and the nonlinear term is truncated at $k_{\rm max}=N/3$. In order to
have a robust method that is also precise, when there is no forcing and
dissipation, we timestep by using a fourth order Runge-Kutta (RK) method.
For weak viscosities, the same RK timestep is used for the deterministic part
(nonlinear and dissipation) and the white noise is added independently, as an
extra (explicit) step.
The timestep has thus to be smaller
than a fraction of $1/(\nu k_{\rm max}^2)$ and $1/(u_{\rm rms}k_{\rm max})$.
For large viscosities, we use instead the implicit method of reference \cite{Mannella1989}.
The noise intensity $D$ is fixed 
by using equation \eqref{eq:Diffufinal}. 
The initial data is set up as a Gaussian white noise in
$x$ with given value of $u_{\rm rms}=\sqrt{2 E}$.
Computations are performed for a number (typically$ N_{\rm rea}=128)$ of independent realisations of the
initial conditions and noise and the results are averaged over realisations. 

\subsection{Physical parameters}

Because of our choice of working with $2\pi$ periodic boundary conditions, 
the largest scale $L$ in our simulations is always fixed to $L= 2 \pi$.
The smallest available scale is resolution dependent and related to the largest wavenumber $k_{\rm max}=[N/3]$ 
(equivalently to the collocation mesh size $\Delta x=2 \pi/N$). 
Thus, a given computation is parametrized by $u_{\rm rms}$, $k_{\rm max}$, and $\nu$.
The initial data used to start the time-integrations is always set to a random Gaussian field 
(see \eqref{eq:pstationary}) with the same value of $u_{\rm rms}$ used to fix
$D$ to its viscosity-dependent value.
Therefore, the $\nu=0$ (and $D=0$) case amounts to integrating the inviscid truncated Burgers equation 
starting from absolute equilibrium initial conditions and, when $\nu$ in non-zero, it is the full 
KPZ system \eqref{eq:KPZ2} (with $\lambda=1$) that is integrated, also starting from the equilibrium distribution.
Thus, in this latter case, one expects to recover the KPZ scaling of the correlation-time in the limit of large spatial scales. 

However, the speed of this approach will depend on the value of the parameters at small scale.
We introduce a scale-dependent Reynolds number:
\begin{equation}
R_e(k)=\frac{u_{\rm rms}}{\nu k}.
 \label{eq:Reynolds}
\end{equation}
The truncation-scale Reynolds number is given by $R_{\rm min}=R_e(k_{\rm max})$, thus
$R_{\rm min}=\frac{u_{\rm rms}}{\nu k_{\rm max}}$, or
\begin{equation}
R_{\rm min}=\frac{3}{N \nu}u_{\rm rms}.
\end{equation}
On general grounds, one expects to see EW scaling when $R_{\rm min}<<1$ and recover the inviscid truncated Burgers case 
in the $R_{\rm min}\to \infty$ limit that corresponds to $\nu=0$.
In what follows, we will determine the time-scale by fixing $u_{\rm rms}=1$ and varying $\nu$. 
We will discuss the crossover in terms of the dimensionless parameter $R_{\rm min}$.
\begin{figure}
\includegraphics[width=0.5\linewidth]{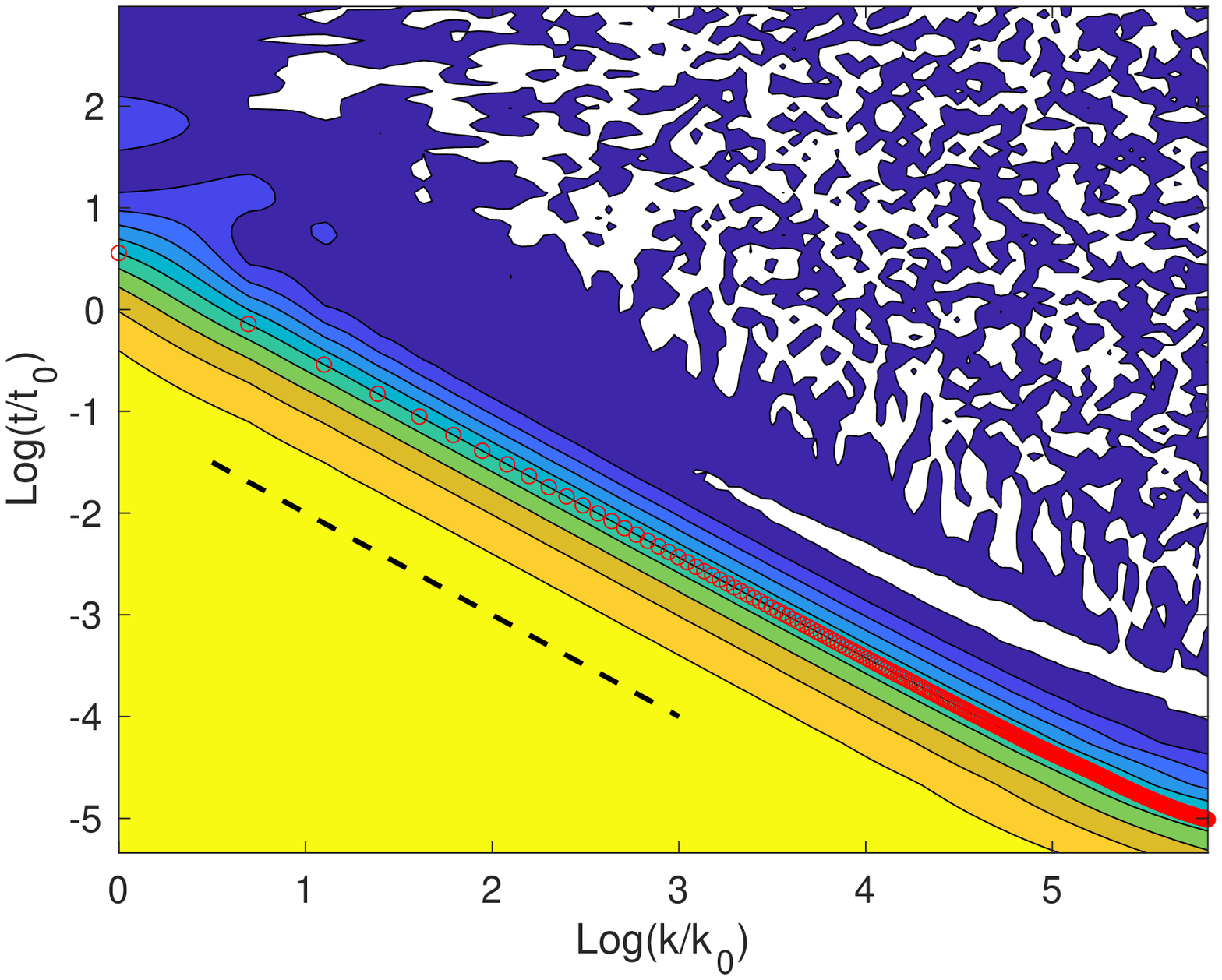}
\includegraphics[width=0.5\linewidth]{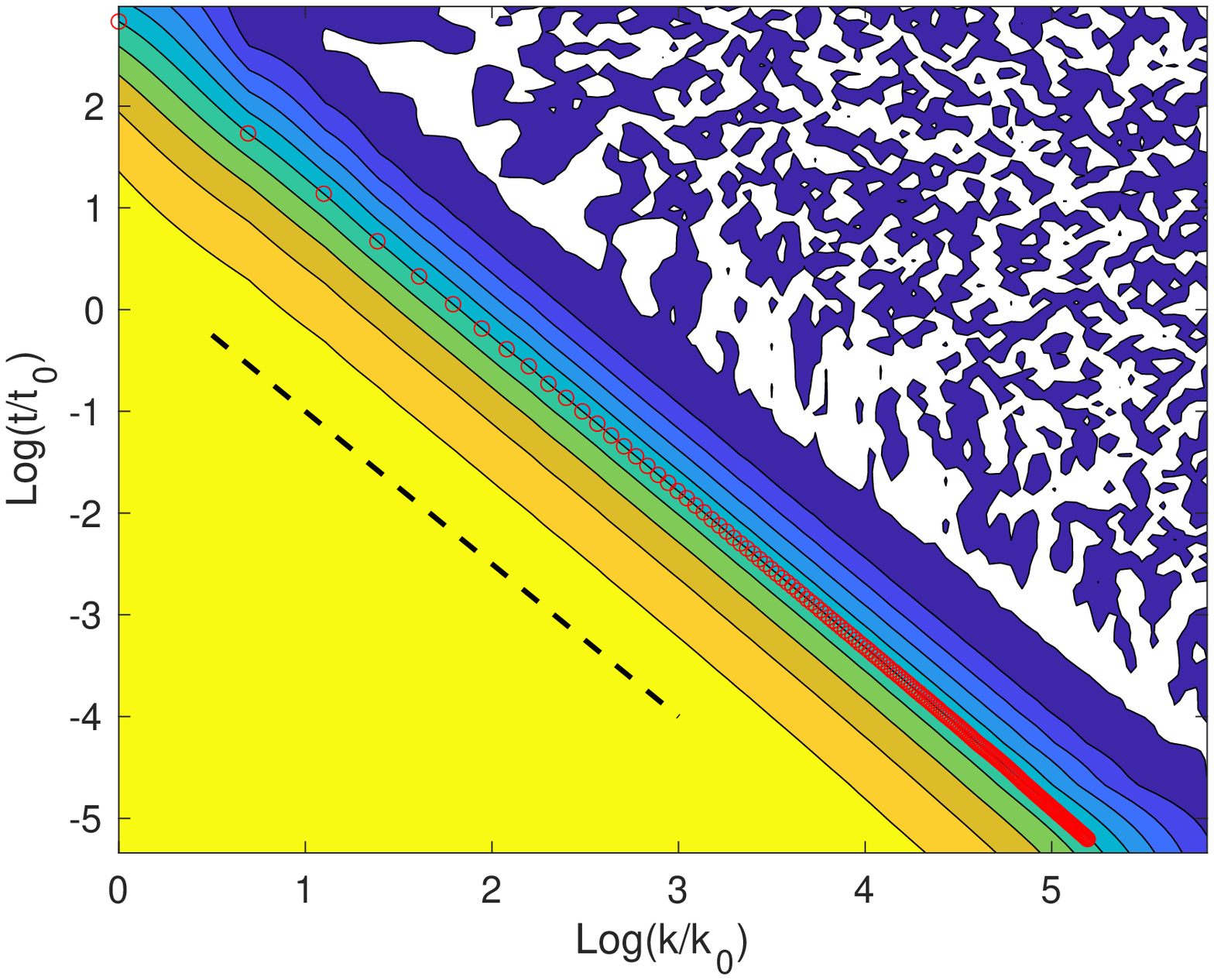}
\caption{
Contour plots of the correlation function $\Gamma(k,t)/\Gamma(k,0)$ represented in the $(\log(k),\log(t))$ plane; (left) Inviscid Burgers scaling obtained at 
$\nu=0$; (right): KPZ scaling obtained at $\nu=3.0 \times10^{-3}$; $\Gamma(k,t)/\Gamma(k,0)$ contour levels are drawn from $0.0$ to $0.9$ and spaced by $0.1$
The dashed black lines indicate (left) $t\sim k^{-1}$ and (right) $t\sim k^{-3/2}$. 
The points $t=\tau_{\frac{1}{2}} (k)$, computed independently using $\Gamma(k,\tau_{\frac{1}{2}})={\frac{1}{2}}\Gamma(k,0)$, 
are indicated by red circles.
}
\label{fig:contours1}
\end{figure}
\begin{figure}
\includegraphics[width=0.5\linewidth]{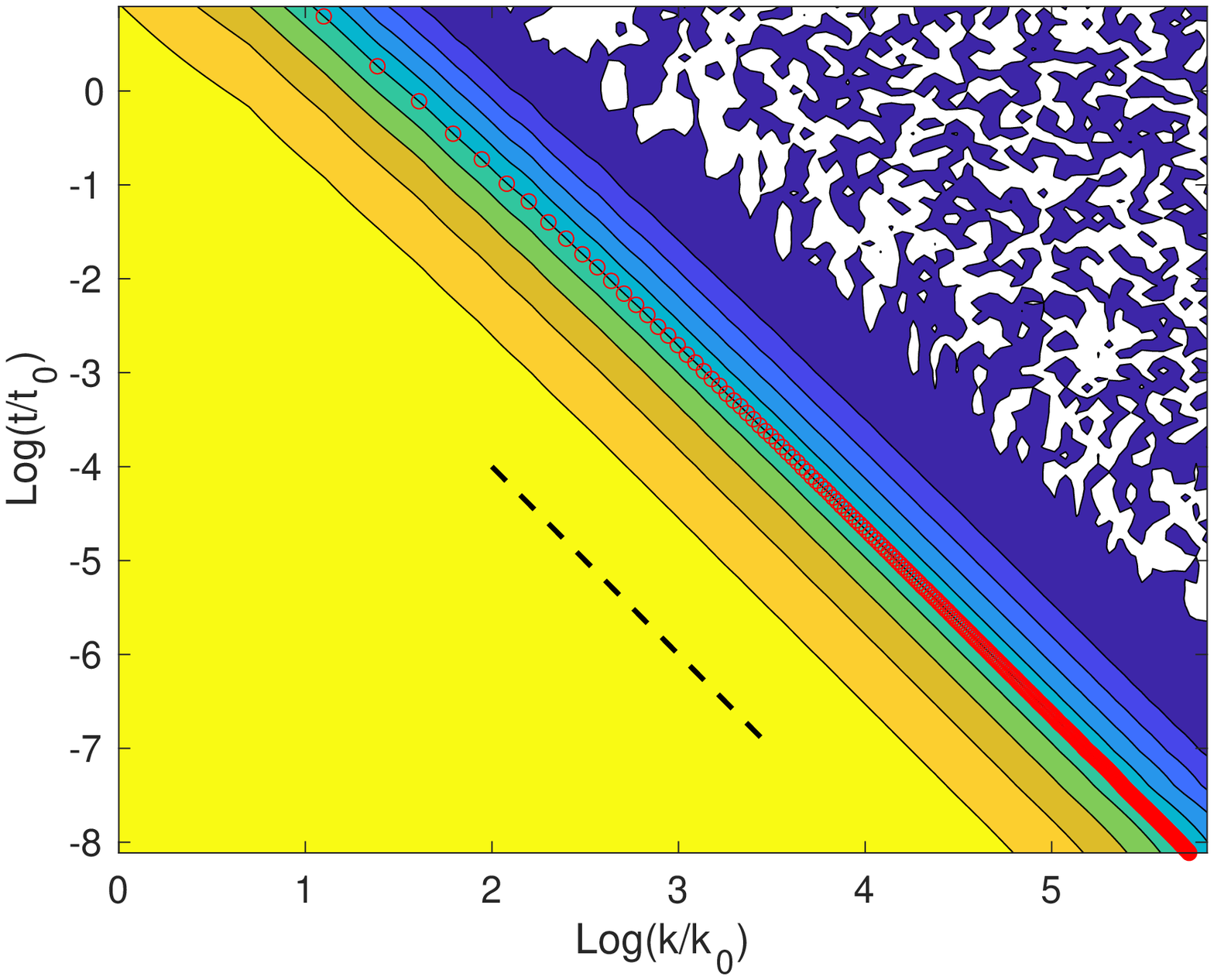}
\includegraphics[width=0.5\linewidth]{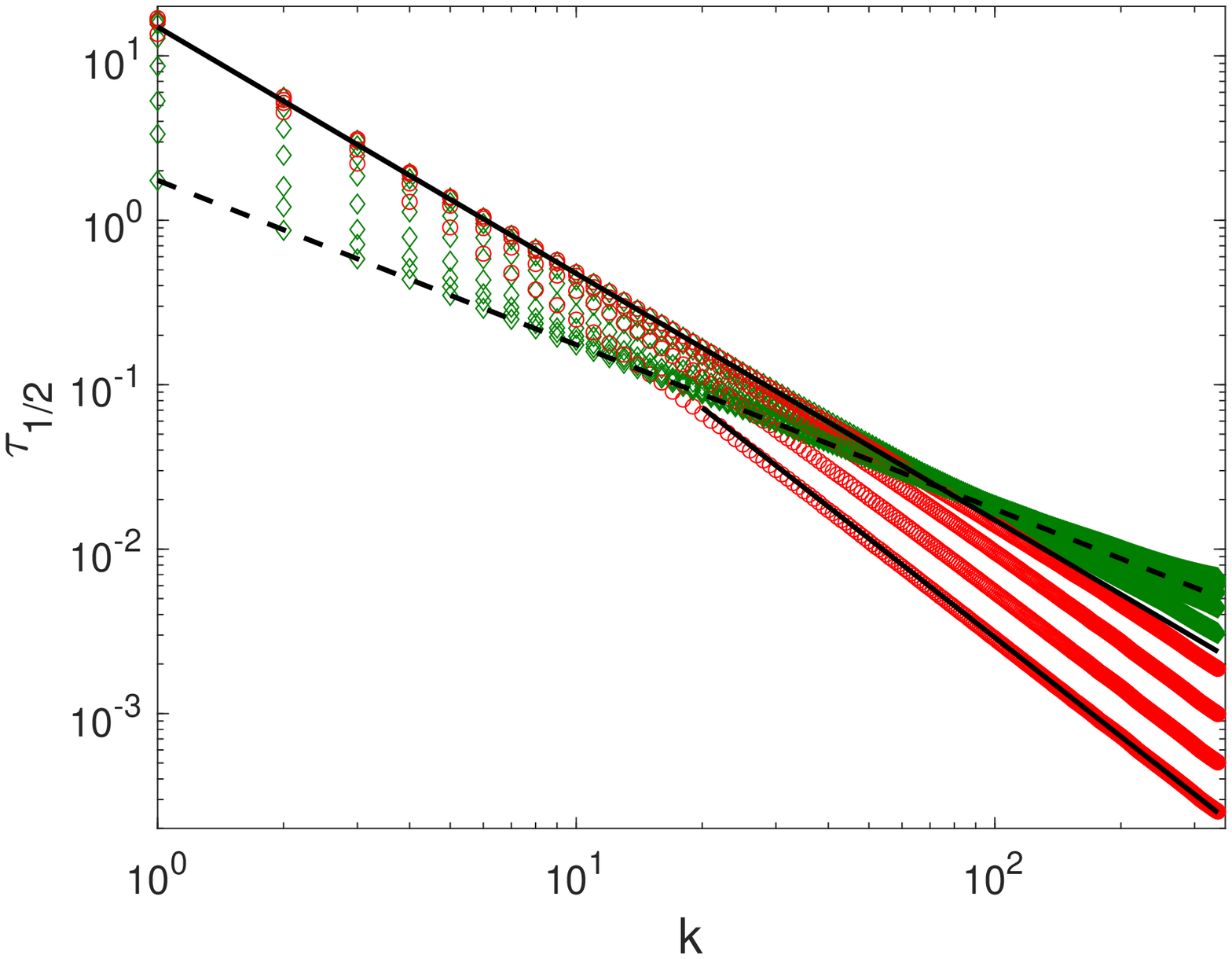}
\caption{
(Left) contour plots of the correlation function $\Gamma(k,t)/\Gamma(k,0)$ represented in the $(\log(k),\log(t))$ plane for the EW scaling  obtained at $\nu=2.4 \times 10^{-2}$; contour levels are drawn from $0.0$ to $0.9$ and spaced by $0.1$.
The dashed black lines indicate $t\sim k^{-2}$. 
The points $t=\tau_{\frac{1}{2}} (k)$, computed independently using $\Gamma(k,\tau_{\frac{1}{2}})={\frac{1}{2}}\Gamma(k,0)$, 
are indicated by red circles. (Right) crossover in the scaling of the decorrelation time $\tau_{\frac{1}{2}}$: 
$\tau_{\frac{1}{2}}$ versus $k$.
Red circles correspond to the EW to KPZ transition:
$\nu=2.4 \times10^{-2}$, $\nu=1.2 \times10^{-2}$,  $\nu=6.0 \times10^{-3}$ and $\nu=3.0 \times10^{-3}$
and green diamond correspond to the KPZ to inviscid transition: 
$\nu=1.5 \times10^{-3}$, $\nu=7.5 \times10^{-4}$,  $\nu=3.8 \times10^{-4}$, 
$\nu=1.9 \times10^{-5}$, $\nu=9.4 \times10^{-6}$, $\nu=4.7 \times10^{-6}$, and $\nu=0$.
Scaling laws are indicated by solid lines: EW $k^{-2}$ and KPZ $k^{-3/2}$. The new inviscid $k^{-1} $ scaling is denoted by a dashed line.
Runs performed with $k_{\rm max}=341$ and 
$u_{rms}=1$.
}
\label{fig:contours2}
\end{figure}
\subsection{Scalings of correlation times}
We first study the behaviour of the the correlation function $\Gamma(k,t)$ by making a series of runs at resolution $N=1024$
($k_{\rm max}=341$), $u_{rms}=1$ and various viscosities.

Scaling behaviour is particularly apparent when represented in the $(\log(t),\log(k))$ plane. 
Indeed, it this logarithmic representation, scaling simply corresponds to equal values of the correlation $\Gamma(k,t)$
along straight lines.

Figure \ref{fig:contours1} shows contour plots of the normalized correlation function 
$\tilde \Gamma(k,t)=\Gamma(k,t)/\Gamma(k,0)$.
The red circles \footnote{Red circles standing right on top of the $0.5$ contour line validate the interpolation scheme that we use to draw the contour lines in the logarithmic representation from the equally spaced in time and wavenumber raw data for $\Gamma(k,t)$.} indicate the points $t=\tau_{\frac{1}{2}} (k)$.
They were computed independently, for each $k$, by solving the equation $\Gamma(k,\tau_{\frac{1}{2}})={\frac{1}{2}}\Gamma(k,0)$.

The left panel of Figure \ref{fig:contours1} shows the inviscid Burgers scaling that is obtained for $\nu=0$, 
corresponding to an infinite truncation-scale Reynolds number $R_{\rm min}$. 
The black solid line indicates the theoretical $t\sim k^{-1}$ law. 
On the right panel, the KPZ scaling obtained at $\nu=3.0 \times10^{-3}$ is displayed, 
corresponding to a truncation-scale Reynolds number $R_{\rm min}=0.98$.
The black solid lines indicating the theoretical $t\sim k^{-3/2}$ law.

The left panel of Figure \ref{fig:contours2} demonstrates the EW scaling that is obtained with $\nu=2.4 \times 10^{-2}$, 
corresponding to a truncation-scale Reynolds number of $R_{\rm min}=0.12$.
The viscous $t\sim k^{-2}$ EZ law is indicated by the black solid line.
The right panel shows the crossover 
in the scaling of the decorrelation times 
$\tau_{\frac{1}{2}}(k)$ versus $k$ for various values of the viscosity $\nu$. 
\begin{figure}
\includegraphics[width=0.8\linewidth]{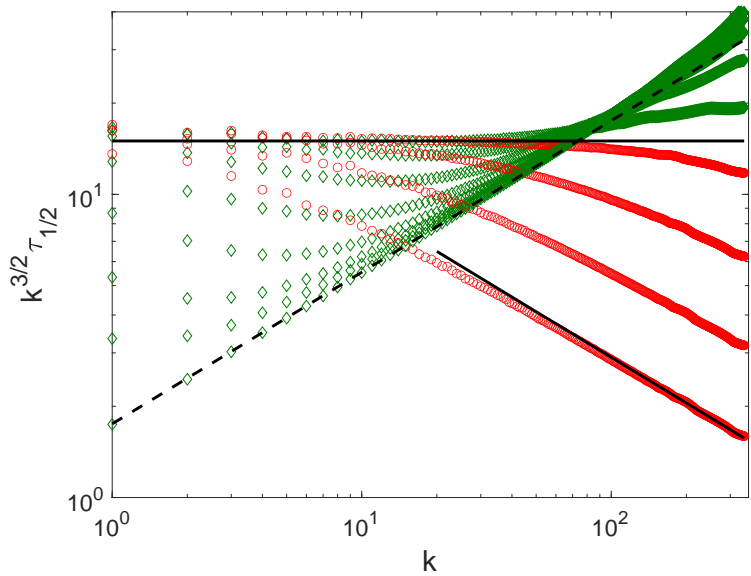}
\caption{
Crossover in the scaling of the decorrelation time $\tau_{\frac{1}{2}}$
for the same conditions as in Figure \ref{fig:contours2} (right panel), but compensated by $k^{3/2}$:
$k^{3/2} \tau_{\frac{1}{2}}$ versus $k$.
Red circles correspond to the EW to KPZ transition:
$\nu=2.4 \times10^{-2}$, $\nu=1.2 \times10^{-2}$,  $\nu=6.0 \times10^{-3}$ and $\nu=3.0 \times10^{-3}$
and green diamond correspond to the KPZ to inviscid transition: 
$\nu=1.5 \times10^{-3}$, $\nu=7.5 \times10^{-4}$,  $\nu=3.8 \times10^{-4}$, 
$\nu=1.9 \times10^{-5}$, $\nu=9.4 \times10^{-6}$, $\nu=4.7 \times10^{-6}$, and $\nu=0$.
Scaling laws are indicated by solid lines: EW $k^{-2}$ and KPZ $k^{-3/2}$. The new inviscid $k^{-1} $ scaling is denoted by a dashed line.
Runs performed with $k_{\rm max}=341$ and 
$u_{rms}=1$.
}\label{fig:crossover1}
\end{figure}
The crossover behaviour is clearly visible in Figure \ref{fig:crossover1} that displays the same data compensated 
by $k^{3/2}$, so that KPZ scaling corresponds to a horizontal line.
The red circles correspond to the EW to KPZ transition with various values of viscosities in geometric progression 
corresponding to a truncation-scale Reynolds number: 
$R_{\rm min}=0.12$, $R_{\rm min}=0.24$, $R_{\rm min}=0.98$.
The green diamond correspond to the KPZ to inviscid transition, with various viscosities, also in geometric progression
corresponding to a truncation-scale Reynolds number: 
$R_{\rm min}=1.96$, $R_{\rm min}=3.91$, $R_{\rm min}=7.71$, $R_{\rm min}=15.4$, $R_{\rm min}=31.2$, $R_{\rm min}=62.3$ and $R_{\rm min}=\infty$  .
The EW $k^{-2}$ scaling law and the KPZ $k^{-3/2}$ scaling are indicated by solid lines and the inviscid $k^{-1} $ scaling is denoted by a dashed line.

\begin{figure}
\includegraphics[width=0.5\linewidth]{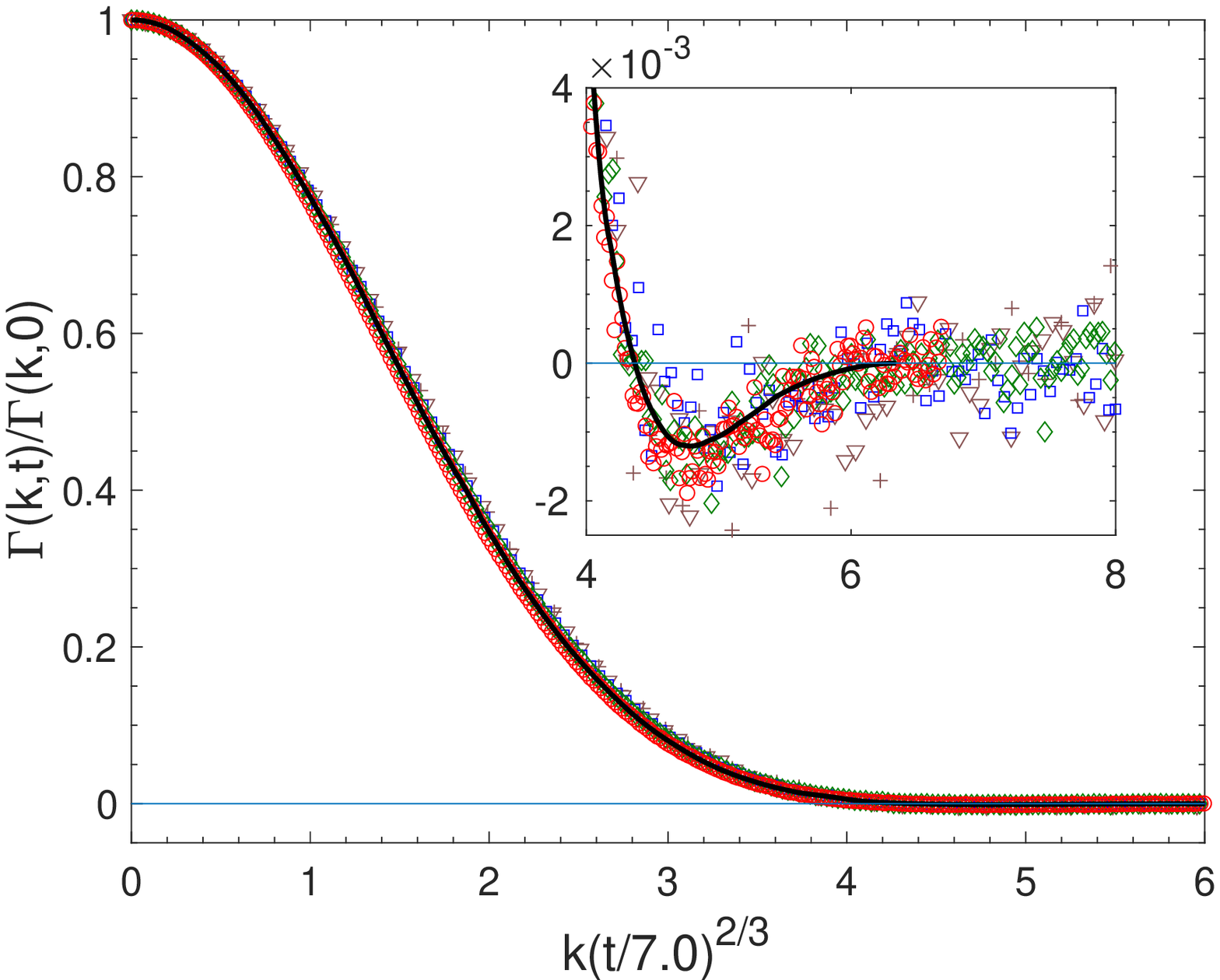}
\includegraphics[width=0.5\linewidth]{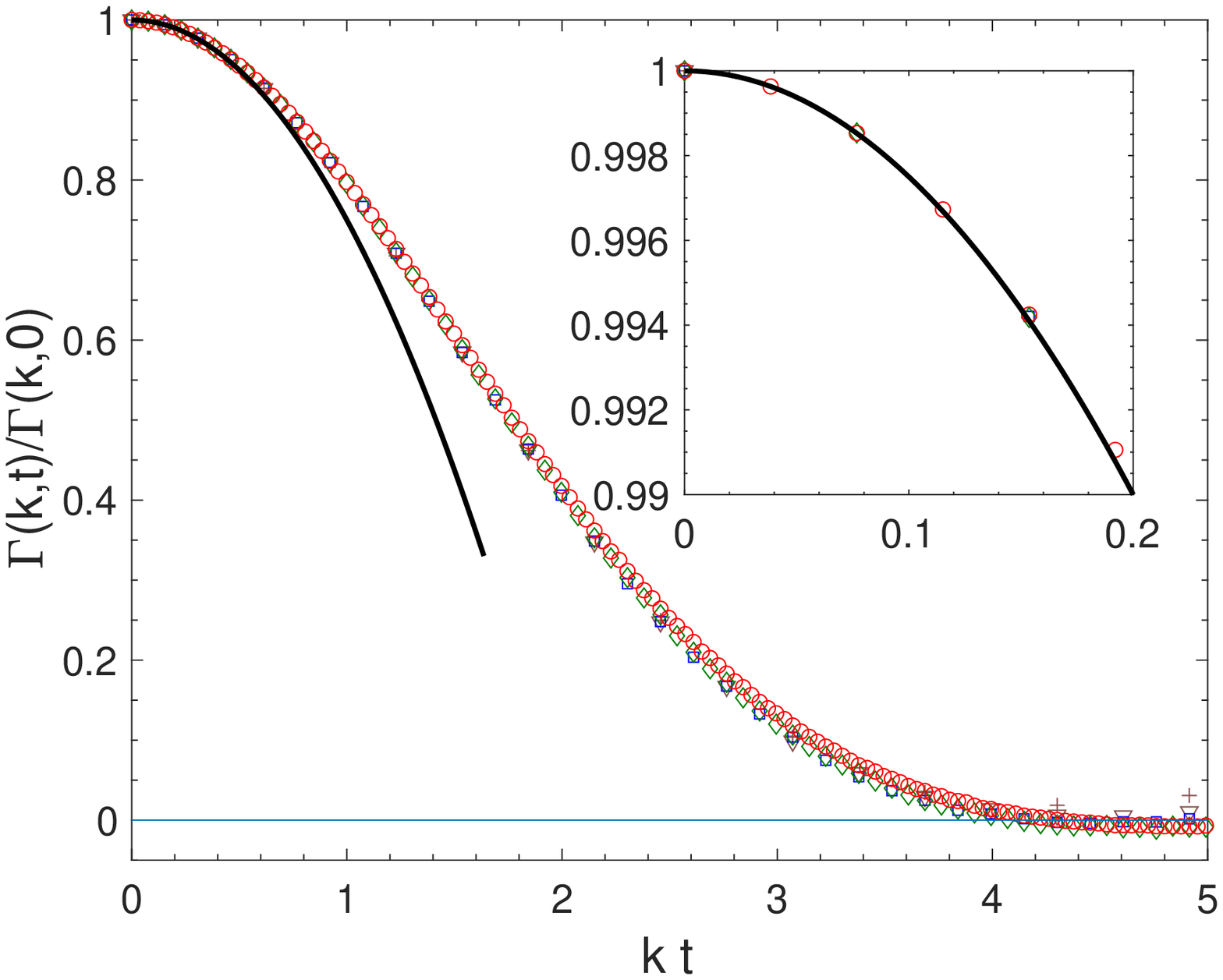}
\caption{Scaling of correlation functions $\Gamma(k,t)/\Gamma(k,0)$ versus the rescaled wavenumber.
Left: KPZ ($\nu=.003$) correlation versus $k (t/7.0)^{2/3}$ (o: $k=4$, diamond: $k=8$, square: $k=16$, v:  $k=32$ and  +: $k=64$) compared with the theoretical correlation function  computed in reference \cite{Spohn2004} shown by a solid line (the inset shows the change of sign of the theoretical correlation); right Inviscid ($\nu=0$) correlation versus $k t$ (o: $k=8$, diamond: $k=16$, square: $k=32$, v:  $k=64$ and  +: $k=128$)  compared with the theoretical short-time parabolic behaviour (see inset and Eq. \eqref{eq:Para}).
Computations were performed with $k_{\rm max}=341$ and $u_{rms}=1$
}\label{fig:rescaled1}
\end{figure}

Figure \ref{fig:rescaled1} shows the scaling form of the normalized correlation functions $\Gamma(k,t)/\Gamma(k,0)$ 
versus the rescaled wavenumber,
with the same conditions as in Fig.\ref{fig:contours1}, Fig.\ref{fig:contours2}  and  Fig.\ref{fig:crossover1}.

The left panel shows the KPZ correlation (obtained at $\nu=.003$), plotted versus the rescaled variable $k (t/7.0)^{2/3}$
for various values of $k$. The theoretical correlation function, computed in reference \cite{Spohn2004}, is shown as a solid black line.
The inset provides details on the change of sign of the correlation. 

The right panel displays the inviscid ($\nu=0$) correlation versus the rescaled variable $k t$, for various values of $k$. 
The theoretical short-time parabolic behaviour is shown as a black continuous curve.
The inset shows short-times details.
The  inviscid parabolic law can be obtained by the following arguments. 
Starting from the equilibrium correlation functions 
$\left<\hat{u} ({k},t) \hat{u} ({k'},0) \right>$,
we can define the time scale $\tau_{\rm C}$ as the parabolic decorrelation time
\begin{equation} 
\tau_{\rm C}^2\partial_{tt} \left<\hat{u} ({k},t) \hat{u} ({k'},0) \right>_{| t=0}  =\left<\hat{u} ({k},0) \hat{u} ({k'},0) \right> ,
\label{taucdef}
\end{equation}
time translation invariance allows us to express the second order time derivative as
\begin{equation} 
-\left<
{\partial_{t}\hat v} ({k},t) \partial_{t'}\hat{u} ({k'},t') \right>_{| t=t'=0}. 
\end{equation}
Using expressions \eqref{eq:NLTrunc2} for the time derivatives
reduces the evaluation of $\tau_{\rm C}$
to that of an equal-time fourth-order moment of a Gaussian field with correlation
$<\hat{u} ({k},t) \hat{u} ({-k},t)>=u_{\rm rms}^2/(2 k_{\rm max}+1)$.
The only non-vanishing contribution is a one loop graph \cite{isserlis,frischbook}.
The correlation time $\tau_{\rm C}$ associated to wavenumber $k$ is found \cite{Cichowlas_Thesis,Cichowlas:2005p1852} 
in this way to obey the simple scaling law
\begin{equation}
\tau_{\rm C}=\frac{\sqrt{2}}{k u_{\rm rms}},
\label{eq:Para}
\end{equation}
and this time-scale is proportional to the eddy turnover time \cite{Majda2000} at wavenumber $k$.

It is apparent from the Figure that our numerical data obey the theoretical predictions.
\subsection{Distributions of the interface increments}
\begin{figure}
\includegraphics[width=0.5\linewidth]{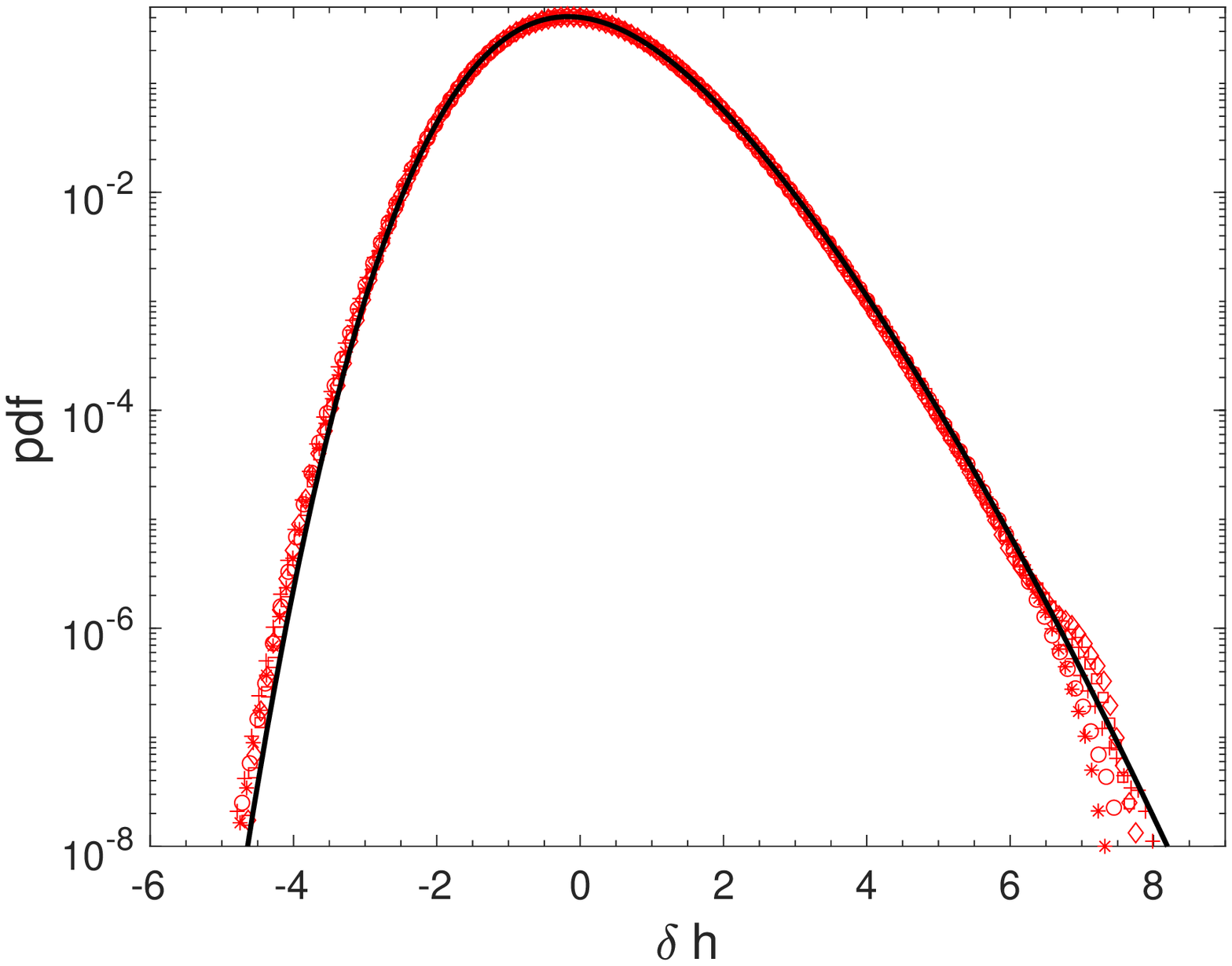}
\includegraphics[width=0.5\linewidth]{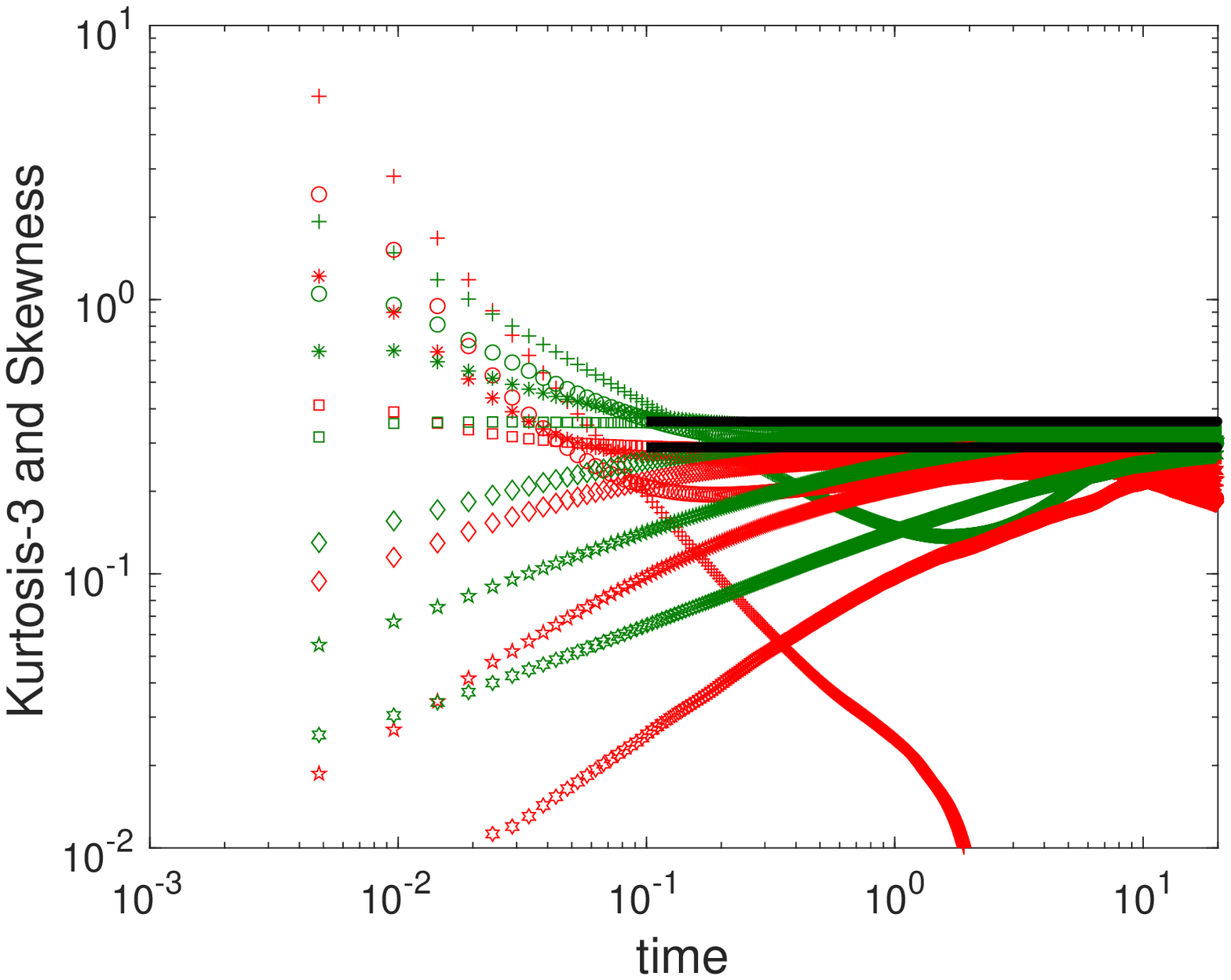}
\caption{
Left: probability distribution functions of the interface increments $\delta h$ at times 
$t=0.5$, $t=1$, $t=2$, $t=4$ and $t=8$, with $\nu=3.0 \times10^{-3}$,
$k_{\rm max}=682$ and $u_{rms}=1$; 
(right) time evolution of the skewness (green) and excess kurtosis (red) at various viscosities in log-log scales.
The solid black lines indicate the theoretical results of reference \cite{Halpin-Healy2014}.
The markers correspond to: 
inviscid  $(\nu=0$): +,
 $\nu=3.8 \times10^{-4}$: o,
 $\nu=7.5 \times10^{-4}$: asterisk,
$\nu=1.5 \times10^{-3}$: square,
$\nu=3.0 \times10^{-3}$: diamond,
$\nu=6.0 \times10^{-3}$: pentagram,
$\nu=1.2 \times10^{-2}$: hexagram;
}\label{fig:pdfskewkurt1}
\end{figure}
The seminal
work of Pr\"{a}hofer and Spohn \cite{prahofer2000} was recently referred to as ``the $2^{nd}$ KPZ
Revolution''~\cite{thhkat2015}.
It has led
to a new set of studies of the 1D KPZ universality
class~\cite{sasamoto2010,calabrese2011,imamura2012,corwin2012,Halpin-Healy2014,quastel2015,saberi-naserabadi-krug-2019}.
In particular, Ref. \cite{Roy2020} notes that, at $x$ and large $t$
\begin{equation}
	\delta h=h(x,t) - h(x,0) \approx  v_{\infty} t + ( \Gamma t)^{\beta_{\text{KPZ}}} \chi_\beta + 
	o(t^{\beta_{\text{KPZ}}}) \ , \ \text{for} \ t \rightarrow \infty,
	\label{eq:KPZh}
\end{equation}
where the parameters $\Gamma$ and $v_\infty$ depend on the model, and $\beta_{\text{KPZ}}=1/3$. 
Furthermore, $\chi_\beta$ is a random variable distributed according to different (Tracy-Widom) 
distributions \cite{tracy1994} 
for different initial conditions. 
For the problem we study, we use Brownian initial data, so we expect 
(see Ref. \cite{Roy2020}) to find a Baik-Rains 
\cite{baik2000} distribution.

Figure \ref{fig:pdfskewkurt1} shows the evolution of the distributions of the interface increments $\delta h$
computed with $k_{\rm max}=682$ and $u_{rms}=1$.
The left panel displays the probability distribution functions of $\delta h$ at various times and  
$\nu=3.0 \times10^{-3}$.
The solid black curve indicates the theoretical probability distribution function of reference \cite{Halpin-Healy2014}.
The right panels show the time evolution of the skewness $S$ (in green) and excess kurtosis $K-3$ (in red) 
at various viscosities. 

\begin{figure}
\includegraphics[width=0.8\linewidth]{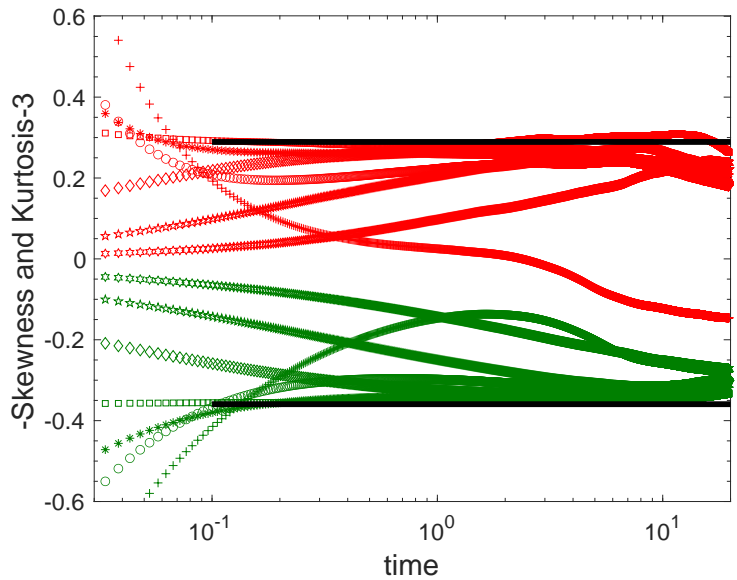}
\caption{
Time evolution of the skewness (green) and excess kurtosis (red) at various viscosities in log-lin scales 
(same conditions as in Figure \ref{fig:pdfskewkurt1} right, with a change of sign for the skewness).
The solid black lines indicate the theoretical results of reference \cite{Halpin-Healy2014}.
The markers correspond to: 
inviscid  $(\nu=0$): +,
 $\nu=3.8 \times10^{-4}$: o,
 $\nu=7.5 \times10^{-4}$: asterisk,
$\nu=1.5 \times10^{-3}$: square,
$\nu=3.0 \times10^{-3}$: diamond,
$\nu=6.0 \times10^{-3}$: pentagram,
$\nu=1.2 \times10^{-2}$: hexagram;
}\label{fig:pdfskewkurt2}
\end{figure}

Figure \ref{fig:pdfskewkurt2} displays the same data, with a change of sign for the skewness (in order to be 
comparable with Figure 3 of Ref. \cite{Roy2020}).
Our results indicate a tendency for the viscous run to converge towards the theoretically predicted values, while the inviscid 
computation only display power law behaviour for the skewness and the excess flatness.
\section{Conclusion} \label{sec:conc}
By using pseudospectral numerical simulations of the $1D$ Galerkin-truncated Burgers equation, 
with noise and dissipation \eqref{eq:KPZ2}, 
we have reproduced the very well known properties  of the $1D$ KPZ universality class
such as the $k^{-3/2}$ scaling of the correlation time, 
the known analytical forms for the rescaled time-correlation function and 
the rescaled interface increments probability distribution function.

We have characterised a new crossover, controlled by the
truncation-scale Reynolds number $R_{\rm min}$, 
towards an inviscid regime with correlation time scaling as $k^{-1}$.
This new regime correspond to the absolute equilibrium solutions of the inviscid noiseless Burgers equation.
To obtain this regime, it is crucial that the numerical scheme {\it exactly} conserves the invariant energy of the system.
This inviscid regime should also be present in finite-difference schemes, provided that they also conserve 
the invariants (see the discussions in Refs. \cite{Majda2000,Majda2002}).
On general grounds, one expects that 
any model within the $1D$ KPZ universality class 
could also exhibit this new crossover
provided that it
also admits {\it exact conservations laws} in some well-defined non-dissipative limit.

This new regime might be amenable to a renormalization-group analysis, 
which would have, in addition to the known \cite{KPZ} KPZ stable fixed point and EW unstable fixed point,
a new fixed point corresponding to the inviscid regime. This is left for a future work. 

\vskip 6pt
\enlargethispage{20pt}
\aucontribute{All authors participated in the analytical computations. MB and RP drafted the manuscript. MB and CC performed the numerical simulations. CC RP and MB read, edited and approved the manuscript. Enrique Tirapegui was instrumental in the early definition of the research project and the corresponding analytical computations. He sadly passed away in 2020 and could not therefore edit and approve to the final manuscript.}
\competing{The authors declare that they have no competing interests.}
\funding{This  work  was  supported  by the French Agence nationale de la recherche (ANR  QUTE-HPC  project  No. ANR-18-CE46-0013).}
\ack{This work was granted access to HPC resources of MesoPSL financed by Region Ile de France and the project Equip@Meso (reference ANR-10-EQPX-29-01) of the  programme  Investissements  d'Avenir  supervised  by Agence Nationale  pour la Recherche. MB and RP thank the Indo-French Centre for Applied Mathematics for financial support. 
RP also thanks CSIR, UGC, and DST India for support and Dipankar Roy for discussions.
CC, ET and MB acknowledge the support of the Laboratoire International Associ\'e "Matie\`re: Structure et Dynamique" LIA-MSD.
CC wishes to acknowledge the support of FONDECYT (CL), No. 1200357 and Universidad de los Andes
(CL) through FAI initiatives.}
\bibliography{biblio} 

\begin{thebibliography}{10}

\bibitem{LEE:1952p4100}
T.~D. Lee, ``On some statistical properties of hydrodynamical and
  magneto-hydrodynamical fields,'' {\em Quart Appl Math}, vol.~10, no.~1,
  pp.~69--74, 1952.

\bibitem{hopf1952statistical}
E.~Hopf, ``Statistical hydromechanics and functional calculus,'' {\em Journal
  of rational Mechanics and Analysis}, vol.~1, pp.~87--123, 1952.

\bibitem{KRAICHNAN:1955p3039}
R.~Kraichnan, ``On the statistical mechanics of an adiabatically compressible
  fluid,'' {\em J. Acoust. Soc. Am.}, vol.~27, no.~3, pp.~438--441, 1955.

\bibitem{KRAICHNARH:1973p2909}
R.~Kraichnan, ``Helical turbulence and absolute equilibrium,'' {\em J. Fluid
  Mech.}, vol.~59, pp.~745--752, 1973.

\bibitem{OrszagHouches}
S.~Orszag, {\em Statistical Theory of Turbulence}.
\newblock in, Les Houches 1973: Fluid dynamics, R.\ Balian and J.L. Peube eds.
  Gordon and Breach, New York, 1977.

\bibitem{Cichowlas:2005p1852}
C.~Cichowlas, P.~Bona{\"\i}ti, F.~Debbasch, and M.~Brachet, ``Effective
  dissipation and turbulence in spectrally truncated {E}uler flows,'' {\em
  Physical Review Letters}, vol.~95, no.~26, p.~264502, 2005.

\bibitem{Krstulovic:2009p1876}
G.~Krstulovic, P.~D. Mininni, M.~E. Brachet, and A.~Pouquet, ``Cascades,
  thermalization, and eddy viscosity in helical {G}alerkin truncated {E}uler
  flows,'' {\em Phys. Rev. E}, pp.~1--5, May 2009.

\bibitem{Got-Ors}
D.~Gottlieb and S.~A. Orszag, {\em Numerical Analysis of Spectral Methods}.
\newblock Philadelphia: SIAM, 1977.

\bibitem{Canuto-Hussaini}
C.~Canuto, M.~Y. Hussani, A.~Quarteroni, and T.~A. Zang, {\em Spectral Methods
  in Fluid Dynamics}.
\newblock New York and Berlin: Springer-Verlag, 2nd printing~ed., 1988.

\bibitem{Krstulovic:2009IJBC}
G.~Krstulovic, C.~Cartes, M.~Brachet, and E.~Tirapegui, ``Generation and
  charecterization of absolute equilibrium of compressible flows,'' {\em
  International Journal of Bifurcation and Chaos}, vol.~19, no.~10,
  pp.~3445--3459, 2009.

\bibitem{Krstulovic2011a}
G.~Krstulovic and M.~E. Brachet, ``Dispersive bottleneck delaying
  thermalization of turbulent {B}ose-{E}instein condensates,'' {\em Phys. Rev.
  Lett}, vol.~106, p.~115303, 2011.

\bibitem{Shukla2013}
V.~Shukla, M.~Brachet, and R.~Pandit, ``Turbulence in the two-dimensional
  {F}ourier-truncated {G}ross-{P}itaevskii equation,'' {\em New J. Phys.},
  vol.~15, no.~11, p.~113025, 2013.

\bibitem{Pouquet2011}
G.~Krstulovic, M.-E. Brachet, and A.~Pouquet, ``Alfv\'en waves and ideal
  two-dimensional {G}alerkin truncated magnetohydrodynamics,'' {\em Phys. Rev.
  E}, vol.~84, p.~016410, Jul 2011.

\bibitem{Ray2011}
S.~S. Ray, U.~Frisch, S.~Nazarenko, and T.~Matsumoto, ``Resonance phenomenon
  for the {G}alerkin-truncated {B}urgers and {E}uler equations,'' {\em Phys.
  Rev. E}, vol.~84, p.~016301, 2011.

\bibitem{Ray2014}
D.~Banerjee and S.~S. Ray, ``Transition from dissipative to conservative
  dynamics in equations of hydrodynamics,'' {\em Phys. Rev. E}, vol.~90,
  p.~041001, Oct 2014.

\bibitem{Ray2015}
S.~S. Ray, ``Thermalized solutions, statistical mechanics and turbulence: An
  overview of some recent results,'' {\em Pramana}, vol.~84, no.~3,
  pp.~395--407, 2015.

\bibitem{Majda2000}
A.~J. Majda and I.~Timofeyev, ``Remarkable statistical behavior for truncated
  {B}urgers-{H}opf dynamics,'' {\em Proceedings of the National Academy of
  Sciences}, vol.~97, no.~23, pp.~12413--12417, 2000.

\bibitem{Majda2002}
A.~Majda and I.~Timofeyev, ``Statistical mechanics for truncations of the
  {B}urgers-{H}opf equation: A model for intrinsic stochastic behavior with
  scaling,'' {\em Milan Journal of Mathematics}, vol.~70, no.~1, pp.~39--96,
  2002.

\bibitem{Cichowlas_Thesis}
C.~Cichowlas, {\em Truncated Euler Equation: from complex singularities
  dynamics to turbulent relaxation}.
\newblock PhD thesis, Universite Pierre et Marie Curie - Paris VI,
  \url{https://tel.archives-ouvertes.fr/tel-00070819/document}, 2005.

\bibitem{FNS}
D.~Forster, D.~R. Nelson, and M.~J. Stephen, ``Large-distance and long-time
  properties of a randomly stirred fluid,'' {\em Phys. Rev. A}, vol.~16,
  pp.~732--749, Aug 1977.

\bibitem{KPZ}
M.~Kardar, G.~Parisi, and Y.-C. Zhang, ``Dynamic scaling of growing
  interfaces,'' {\em Phys. Rev. Lett.}, vol.~56, pp.~889--892, Mar 1986.

\bibitem{thhzhang1995}
T.~Halpin-Healy and Y.-C. Zhang, ``Kinetic roughening phenomena, stochastic
  growth, directed polymers and all that. aspects of multidisciplinary
  statistical mechanics,'' {\em Physics Reports}, vol.~254, no.~4, pp.~215 --
  414, 1995.

\bibitem{thhkat2015}
T.~Halpin-Healy and K.~A. Takeuchi, ``A {KPZ} cocktail-shaken, not
  stirred...,'' {\em Journal of Statistical Physics}, vol.~160, pp.~794--814,
  Aug 2015.

\bibitem{quastel2015}
J.~Quastel and H.~Spohn, ``The one-dimensional {KPZ} equation and its
  universality class,'' {\em Journal of Statistical Physics}, vol.~160,
  pp.~965--984, Aug 2015.

\bibitem{EW1982}
S.~F. Edwards and D.~R. Wilkinson, ``The surface statistics of a granular
  aggregate,'' {\em Proceedings of the Royal Society of London. A. Mathematical
  and Physical Sciences}, vol.~381, no.~1780, pp.~17--31, 1982.

\bibitem{Halpin-Healy2014}
T.~Halpin-Healy and Y.~Lin, ``Universal aspects of curved, flat, and
  stationary-state {K}ardar-{P}arisi-{Z}hang statistics,'' {\em Phys. Rev. E},
  vol.~89, p.~010103, Jan 2014.

\bibitem{frischbook}
U.~Frisch, {\em Turbulence: The Legacy of A. N. Kolmogorov}.
\newblock Cambridge University Press, 1995.

\bibitem{Frisch2001}
U.~Frisch and J.~Bec, ``Burgulence,'' in {\em New trends in turbulence
  Turbulence: nouveaux aspects: 31 July -- 1 September 2000} (M.~Lesieur,
  A.~Yaglom, and F.~David, eds.), (Berlin, Heidelberg), pp.~341--383, Springer
  Berlin Heidelberg, 2001.

\bibitem{BecKhaninPhysRep}
J.~Bec and K.~Khanin, ``Burgers turbulence,'' {\em Physics Reports}, vol.~447,
  no.~1, pp.~1--66, 2007.

\bibitem{Majda2003}
R.~V. Abramov, G.~Kova{\v c}i{\v c}, and A.~J. Majda, ``Hamiltonian structure
  and statistically relevant conserved quantities for the truncated
  {B}urgers-{H}opf equation,'' {\em Communications on Pure and Applied
  Mathematics}, vol.~56, no.~1, pp.~1--46, 2003.

\bibitem{Van-Kampen-Chem}
N.~G.~v. Kampen, {\em Stochastic processes in physics and chemistry}.
\newblock North-Holland ; sole distributors for the USA and Canada, Elsevier
  North-Holland, Amsterdam ; New York : New York :, 1981.

\bibitem{LivreVertET}
F.~Langouche, D.~Roekaerts, and E.~Tirapegui, {\em Functional integration and
  semiclassical expansions}.
\newblock D Reidel Pub Co, Jan 1982.

\bibitem{KRA59}
R.~Kraichnan, ``Classical fluctuation-relaxation theorem,'' {\em Phys. Rev.},
  vol.~113, no.~5, pp.~1181,1182, 1959.

\bibitem{Mannella1989}
V.~Mannella, R.and~Palleschi, ``Fast and precise algorithm for computer
  simulation of stochastic differential equations,'' {\em Phys.Rev.A}, vol.~40,
  pp.~3381-- 3386, Sep 1989.

\bibitem{Spohn2004}
M.~Pr{\"a}hofer and H.~Spohn, ``Exact scaling functions for one-dimensional
  stationary {KPZ} growth,'' {\em Journal of Statistical Physics}, vol.~115,
  pp.~255--279, Apr 2004.

\bibitem{isserlis}
L.~Isserlis, ``On a formula for the product-moment coefficient in any number of
  variables,'' {\em Biometrika}, vol.~12, pp.~134,139, 1918.

\bibitem{prahofer2000}
M.~Pr\"ahofer and H.~Spohn, ``Universal distributions for growth processes in
  $1+1$ dimensions and random matrices,'' {\em Phys. Rev. Lett.}, vol.~84,
  pp.~4882--4885, May 2000.

\bibitem{sasamoto2010}
T.~Sasamoto and H.~Spohn, ``One-dimensional {K}ardar-{P}arisi-{Z}hang equation:
  An exact solution and its universality,'' {\em Phys. Rev. Lett.}, vol.~104,
  p.~230602, Jun 2010.

\bibitem{calabrese2011}
P.~Calabrese and P.~Le~Doussal, ``Exact solution for the
  {K}ardar-{P}arisi-{Z}hang equation with flat initial conditions,'' {\em Phys.
  Rev. Lett.}, vol.~106, p.~250603, Jun 2011.

\bibitem{imamura2012}
T.~Imamura and T.~Sasamoto, ``Exact solution for the stationary
  {K}ardar-{P}arisi-{Z}hang equation,'' {\em Phys. Rev. Lett.}, vol.~108,
  p.~190603, May 2012.

\bibitem{corwin2012}
I.~Corwin, ``The {K}ardar-{P}arisi-{Z}hang equation and universality class,''
  {\em Random Matrices: Theory and Applications}, vol.~01, no.~01, p.~1130001,
  2012.

\bibitem{saberi-naserabadi-krug-2019}
A.~A. Saberi, H.~Dashti-Naserabadi, and J.~Krug, ``Competing universalities in
  {K}ardar-{P}arisi-{Z}hang growth models,'' {\em Phys. Rev. Lett.}, vol.~122,
  p.~040605, Jan 2019.

\bibitem{Roy2020}
D.~Roy and R.~Pandit, ``One-dimensional {K}ardar-{P}arisi-{Z}hang and
  {K}uramoto-{S}ivashinsky universality class: Limit distributions,'' {\em
  Phys. Rev. E}, vol.~101, p.~030103, Mar 2020.

\bibitem{tracy1994}
C.~A. Tracy and H.~Widom, ``Level-spacing distributions and the {A}iry
  kernel,'' {\em Communications in Mathematical Physics}, vol.~159,
  pp.~151--174, Jan 1994.

\bibitem{baik2000}
J.~Baik and E.~M. Rains, ``Limiting distributions for a polynuclear growth
  model with external sources,'' {\em Journal of Statistical Physics},
  vol.~100, pp.~523--541, Aug 2000.

\end{thebibliography}
\bibliographystyle{ieeetr}
%
%
%
\end{document}